\documentclass[manuscript,screen]{acmart}
\AtBeginDocument{%
  }

\setcopyright{acmlicensed}
\copyrightyear{2026}
\acmYear{2026}
\acmDOI{XXXXXXX.XXXXXXX}
\acmISBN{978-1-4503-XXXX-X/2018/06}

\usepackage{caption}
\usepackage{subcaption}
\usepackage{verbatim}

\newif\ifanonymous
\anonymousfalse

\begin{document}

\title{Opportunities of Touch-Enabled Spherical Displays to support Climate Conversations}


\ifanonymous
\else

\author{Mathis Brossier}
\email{mathis.brossier@liu.se}
\orcid{0000-0001-7653-9457}
\affiliation{%
  \institution{Linköping University}
  \country{Sweden}
}

\author{Mina Mani}
\email{mina.mani@liu.se}
\orcid{0000-0003-2669-8059}
\affiliation{%
  \institution{Linköping University}
  \country{Sweden}
}

\author{Agathe Malbet}
\email{agathe.malbet@liu.se}
\orcid{0009-0005-4289-0309}
\affiliation{%
  \institution{Linköping University}
  \country{Sweden}
}

\author{Konrad Schönborn}
\email{konrad.schonborn@liu.se}
\orcid{0000-0001-8888-6843}
\affiliation{%
  \institution{Linköping University}
  \country{Sweden}
}

\author{Lonni Besançon}
\email{lonni.besancon@liu.se}
\orcid{0000-0002-7207-1276}
\affiliation{%
  \institution{Linköping University}
  \country{Sweden}
}

\fi

\renewcommand{\shortauthors}{Brossier et al.}

\begin{abstract}
We explore how touch-sensitive spherical displays can support climate conversations in museums and science centers. These displays enable intuitive and embodied interaction with complex climate data, and support collective exploration. However, current interaction capabilities of spherical displays are limited. Therefore, this exploratory study aims to identify potential opportunities to develop meaningful interactions and technical solutions. Through two workshops, key opportunities were identified to improve visitors’ understanding and navigation of climate data, along with recommendations for technical implementation. Our results provide guidelines and aspects to consider for future research and development in this area.\end{abstract}



\received{20 February 2007}
\received[revised]{12 March 2009}
\received[accepted]{5 June 2009}

\maketitle

\section{Introduction}

Climate change is arguably our most urgent global challenge and requires informed decision-making at both the individual and community levels. Climate data involve huge amounts of information, ranging from temperature anomalies to climate predictions, which are not easily understood. Visualizations can transform complex climate-related data into visual narratives that are more accessible to various audiences, making learning activities about climate change more accessible and understandable \cite{nocke_visualization_2008}. More specifically, spherical displays (interactive digital globes) enable intuitive exploration of planetary data and improve spatial understanding \cite{wang_capturing_2019}. These devices also allow simultaneous and co-located participation of multiple users, contributing to collaborative learning and dialogue among citizens.

Museums and science centers, important public learning spaces \cite{Schwan:2014:UAE} also contribute to bringing science learning and communication to broader communities \cite{Archer:2016:BMH}. When using such devices, they play an essential role in facilitating and stimulating climate-related conversations among visitors \cite{chen_reaching_2020}. However, spherical displays, while popular, represent a very peculiar interaction context. They rarely provide direct touch-based interaction (\cite{soni_towards_2019} despite the technological possibility to do so. Instead, they mostly incorporate non-interactive or indirect-interactive formats, such as pre-programmed visualizations or external flat-screen controls. Furthermore, the existing literature is limited in how these technologies can be leveraged to involve multiple users in public settings \cite{soni_towards_2019,soni_adults_2020,vega_visualization_2014}. Therefore, there is a need to explore meaningful and intuitive direct interactions with spherical displays that aim to facilitate climate conversations and support learning in museum settings.
In response, we present preliminary findings on interaction with a touch-enabled spherical display, focusing on possible gestural patterns and triggers for climate-related conversations. The following research questions guided this exploration: 

\begin{itemize}
    \item How would participants use the spherical display during conversations about climate issues?  
    \item What are the fundamental interaction patterns needed for spherical displays to support climate conversations? 
\end{itemize}

\section{Related work}

We survey examples of previous work on spherical display interaction as well as climate conversations.

\paragraph{Interacting with spherical interfaces} Spherical displays are novel display devices that allow realistic visualization of planet-scale datasets without requiring a projection, contrary to 2D displays, thereby increasing realism. As novel and tangible devices, they increase engagement and stimulate curiosity, making them particularly suited to the context of public spaces \cite{schuman_ocean_2022}. Research on spherical display visualization was pioneered simultaneously by \citeauthor{hruby_global_2008}\cite{hruby_global_2008} and \citeauthor{benko_sphere_2008}\cite{benko_sphere_2008}. They identified two main purposes and research avenues, namely using spherical displays as a communication medium for climate change; and as a novel interactive device. At the same time, \citeauthor{cartwright_digital_2007}\cite{cartwright_digital_2007} studied touch-based interactions on spherical displays. Later, scholars expanded both research directions in public science outreach and interactive visualization. For example,\citeauthor{vega_visualization_2014}\cite{vega_visualization_2014} explored visualization on spherical displays in the context of public spaces and education, while \citeauthor{williamson_globalfestival_2015}\cite{williamson_globalfestival_2015} studied in-situ touch interaction in public spaces. However, because of their recent emergence, spherical devices have not been studied thoroughly in previous research. They offer potential for novel interaction patterns in HCI, such as non-planar surfaces and touch gestures, managing physical points of view and occlusions, as well as collaborative interaction in shared spaces \cite{soni_towards_2019,vega_visualization_2014}. 


\paragraph{Holding climate conversations}

The \textit{climate spiral of silence} is a phenomenon that tends to suppress open discourse on  environmental issues \cite{maibach_is_2016,noelle-neumann_spiral_1993,geiger_climate_2016}. Furthermore, climate obstruction reinforces inaction \cite{ekberg_climate_2022} and can exacerbate climate anxiety, leading people to internalize their fears and anxieties \cite{hickman_climate_2021}. 
Climate conversations are a tool centered around dialogue, which can increase understanding and acceptance of climate change \cite{goldberg_discussing_2019}, and reach individuals who are usually not exposed to climate information \cite{mycoo_communicating_2015}. Value-centered conversations have been found to be more rewarding than didactic teaching alone\cite{hawkins_pause_2014}, and found to result in positive outcomes \cite{bloomfield_effects_2020}, an increase in support for climate policies \cite{geiger_examination_2022} and more effective than conversations purely centered on climate change communication. Researchers often use the deficit model (one-way communication) to present their results. However, dialogue present in climate conversations can increase understanding and acceptance of climate change \cite{goldberg_discussing_2019}, reaching individuals who are not usually exposed to climate information \cite{mycoo_communicating_2015}. Climate conversations, through climate forums and citizen assemblies, lead to more public engagement \cite{devaney_irelands_2020,hanson_public_2018,leal_filho_community_2017}. Such public involvement in climate change encourages collective adaptation \cite{hugel_public_2020} and drives policy support for sustainable infrastructure \cite{perlaviciute_at_2018}. 

Engaging in climate conversations is increasingly crucial because the phenomenon of a climate spiral of silence tends to suppress open discourse on environmental issues \cite{maibach_is_2016,noelle-neumann_spiral_1993,geiger_climate_2016}. This phenomenon explains that the less people hear about contentious issues such as climate change, the less they feel it is acceptable to talk about it. In addition, this climate spiral of silence allows the emergence of alternative narratives such as denial of evidence and denial of response, leading to climate obstruction and reinforcing inaction \cite{ekberg_climate_2022}. The spiral of silence can also exacerbate climate anxiety by preventing open discussions about climate concerns, leading individuals to internalize their fears and anxieties \cite{hickman_climate_2021}. 

In order to break the climate spiral of silence, spaces for discussion must be created \cite{geiger_creating_2017}. Creating such a space involves encouraging and guiding conversations, developing forums to freely exchange ideas, challenging prevailing narratives, and confronting cognitive dissonance related to climate change. The quality and frequency of climate change conversations can be improved through communication training \cite{swim_social_2018}. Group discussions built on shared values have been shown to produce positive outcomes \cite{bloomfield_effects_2020} and increase support for climate policies \cite{geiger_examination_2022}. Furthermore, conversations centered on values have been found to be more rewarding than didactic teaching \cite{hawkins_pause_2014} and more effective than conversations centered purely on a climate change communication \cite{van_swol_fostering_2022}.  

Multiple examples of such conversations already exist, such as the Talk Climate Change campaign that encouraged citizens to hold 26,000 climate conversations with friends, families, and colleagues before COP26 \cite{ettinger_breaking_2023}.  

\section{Method}

\begin{figure}
\centering
\includegraphics[width=\linewidth]{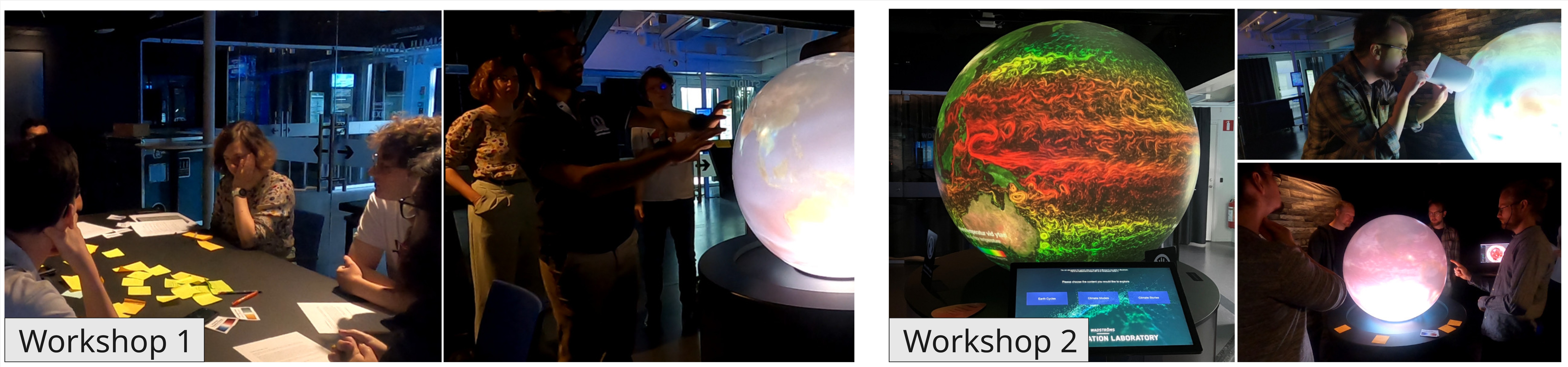}
\caption{Pictures from the two workshops}
\label{fig:workshops}

\end{figure}


We conducted two exploratory workshops at \ifanonymous\else Visualization Center C in Norrköping, Sweden \fi,  to explore potential uses of spherical displays to support climate conversations among groups. We relied on the spherical implementation described in Besançon et al.'s public climate visualization article \cite{besanccon2022exploring}. The first workshop was designed by consulting literature on climate conversations inspired by \citeauthor{ettinger_breaking_2023}\cite{ettinger_breaking_2023} in addition to previous research on spherical display interaction. The second workshop design was informed by results of the first. 

\paragraph{Participants.}

Seven participants were recruited through a convenience sampling approach \cite{robson_real_2016}. They were divided into two respective novice and expert groups, based on their experience with both spherical displays and climate data visualizations. The three participants in Workshop 1 either worked in or studied computer science and the four participants in Workshop 2 worked in communication, sustainability, education, and/or media technology for science centers. We obtained written consent from all participants to take part in the study and for us to use their materials, photos, and recordings. 

\paragraph{Workshop materials.}

Two sets of cards were created to scaffold the workshop: six climate conversation cards with a \emph{conversation starter}, sourced from the Talk Climate Change project \cite{ettinger_breaking_2023}, and selected according to their relevance, as well as four climate visualization cards that allowed participants to select from air surface temperature change, precipitation change, sea ice concentration, and sea surface temperature change. We used two spherical displays: a smaller 60cm touch-sensitive display with two possible interactions—slide to rotate and tap to select—, and a larger 120 cm display controlled with an external fixed touch screen providing descriptions, dataset selection, rotation with swipes, and animation playback control.

\paragraph{Design of workshops}
Both workshops commenced with holding a climate conversation on one of the existing spherical displays. Participants and facilitators (authors) brainstormed on sticky notes (\autoref{fig:workshops}) and conceptualized them on the spherical displays as possible features that could improve the supporting role of spherical displays. 

\paragraph{Data collection and analysis}
Both workshops were video recorded and supplemented with photographs and collected sticky notes. We analyzed the materials reflexively \cite{alvesson_reflexive_2012} by individually identifying emerging themes, then jointly discussing and clustering them in multiple iterations to capture conversational themes, challenges, and opportunities.

\section{Identified interaction opportunities}

Our analysis revealed two clusters of interaction opportunities for spherical displays pertaining to climate data.

\subsection{Understanding the data}

Given the complexity of climate data, \emph{understanding data} is crucial for initiating and carrying out conversations. Participants frequently discussed the meaning of the data during both workshops focusing on the following aspects.

\paragraph{Accessing the legend and the description of the data visualization.}

The legend is essential to accurately interpret symbols, colors, and patterns used. A textual description can also inform users about data origin, limitations and potential biases of the visualization. In Workshop 2, participants discussed the ``suitability for life'' dataset comparing 2020 and 2070. They expressed the uncertainty of their interpretation, unsure about the meaning of colors and potential limitations of the model. Those doubts were cleared after reaching for the legend and description. 

\paragraph{Controlling the amount of information.}
In multiple instances, the participants desired to add or remove layers of information to better make sense of the complex interconnected climate change data. For example, participants were interested in displaying country borders to situate themselves or to combine multiple data visualizations to pinpoint the most suitable places to live in future scenarios. 

\paragraph{Making the data personal.}

Personal experiences such as ecological or temperature changes in participants' local environment can serve as an entry point to interact with climate data and trigger climate conversations. During both workshops, participants referred to their personal lived experiences to make sense of the data visualizations.

\paragraph{Linking external information.}

Similarly, participants exhibited interest in connecting the presented data visualization to external information, such as linking the seasonal surface temperature changes to the climate projections of surface temperature. External information can help participants relate the visualized data to information they already know. Implementation could take multiple forms, including the introduction of additional (localized) information or enabling comparisons to highlight the differences.
\paragraph{Data-visual mapping for designing data visualization}

Interactions can be affected by the design choices made in the creation of the visualizations, such as a range of colors that can mislead users or the stacking of layers of data on a geographical map. A map outline seems to be preferable to a realistic display of the terrain. The current prototype overlays data visualizations on a colored world map with partial opacity, which leads to identical hues appearing differently depending on the terrain.  This caused an issue where the same shade of red appeared darker when overlaid on forest (dark green) than on desert (light yellow) and rendered comparative analysis difficult. This highlights the importance of considering background-foreground interactions and consistency in the design of spatial data visualizations.

\subsection{Navigating the data}

Another aspect to consider in order to understand the data is to be able to navigate it. Navigation in this sense means obtaining granular information, being able to select data visualizations and time ranges within animated visualizations, and comparing information. 

\paragraph{Getting information for specific data points.}

Our observations and the participants' requests underscore the importance of multi-scale data availability. One participant in workshop 2 was wondering why Norway appeared red in the `Where can we live in 2050?' visualization. After noticing that the Alps were also red (not suitable for life) and that the visualization of the data was based on expected precipitation and temperature, the participants hypothesized that mountain ranges might have similar characteristics, such as a lack of rain and poor suitability for agriculture. Participants in Workshop 1 similarly expressed the need for granularity in the presented visualization. They proposed that zooming in, access to raw data of a specific point (e.g., atmospheric $\mathrm{CO_2}$ concentrations, localized temperature anomalies, etc.) or contextualized interpretations, such as the implications of an average +2°C temperature rise in a specific location, would enhance their understanding.  

\paragraph{Controlling the timeline.}

All participants emphasized the importance of controlling animated data visualizations with play/pause buttons and a time scrubbing bar. Such controls facilitate detailed analysis by allowing users to selectively observe specific moments or intervals within the temporal evolution of the data. 


\paragraph{Positioning around the globe and group dynamics}

The spherical display provides a comprehensive, all-encompassing visualization of the globe, leveraging its three-dimensional form to enhance geographic likeness. However, this perceptual advantage entails a limitation: the entire globe cannot be assimilated simultaneously by a single viewer due to physical and perceptual constraints. To mitigate this, users can manipulate the globe through rotation, executed via intuitive touch gestures such as swiping, or by physically repositioning themselves around the display to obtain different viewpoints. From an interaction design perspective, rotating the globe from a fixed point of view offers advantages in terms of operational simplicity and consistency, particularly facilitating accessibility. It also aligns with standard touchscreen navigation. However, when users are dispersed around the spherical display, the rotation of the globe by one individual changes the shared visualization for all, potentially leading to conflicts. Alternatively, regrouping on a single side reduces collaboration, as everyone may look at one single area instead of multiple viewpoints. 

\paragraph{Comparing}

The above limitations of spherical displays also hampers immediate comparative analysis since, for instance, users must walk around the display to compare distant regions. Similarly, climate visualizations often span long timelines. For example, moving from Sweden’s 2020 data to 2050 makes it difficult to recall previous views and compare future conditions. Furthermore, a participant wanted to compare across multiple data visualizations, in order to connect different information (e.g., sea ice temperature and precipitations). Some participants suggested dividing the sphere into pieces with different data visualizations or layering them. Participants in both workshops noted that while patterns can be discerned, working memory constraints make comparisons treacherous. Therefore, tools that enable comparison across the display (spatial), across time (temporal), and across data visualizations need to be further investigated and developed. 

\section{Technical suggestions}

We propose technical implementations for future work on spherical displays that build on our previous analyses.

\paragraph{Map layers.}

Layering data visualizations can address the expressed need to change which data visualizations are made salient. It aims to control the amount of information displayed and to facilitate comparisons with an area by showing or hiding them across different layers. However, layering can create visual clutter and lead to confusion \cite{munzner_visualization_2014}. While distinguishing between oceans and continents is important for orientation, we recommend that distinct color and/or style encodings be reserved for data visualizations. Modulating color values, outlining, or hatching can be explored alongside \cite{he_encoding_2024}. We also recommend that only a limited number of layers be displayed at a time to improve readability, following previous recommendations \cite{munzner_visualization_2014}.

\paragraph{Touch gestures.}

Gestures (panning, rotating, and swiping) allow  comparison of different areas of the globe or control of the timeline. Due to their widespread use in map software, these interactions form a very intuitive and effective way to navigate. 
On a sphere, the pan gesture allows visitors to spin the globe around the equator to see what lies on the other side. This gesture also enables the visitors to tilt the globe up and down to focus on the poles.
As brought up by participants in Workshop 2, unrestrained rotation can leave the globe in a confusing state, such as having the south pole at the top. This can be mitigated by a rotation axis limit, often employed in visualization \cite{besancon:hal-01372922,besançon2020maps}, or a spring animation that returns the globe to the upright position after a delay \cite{williamson_globalfestival_2015}. 
Rotation can also be a source of conflict when multiple people use the spherical display at the same time, as it affects the whole globe at once. Pairing this functionality with physical or digital props may encourage visitors to take turns panning the globe to avoid such conflicts \cite{gennari2020turntalk}. Another approach could be to deactivate panning and encourage people to move around the globe to view different points of interest, but this may increase frustration.
It was also suggested that using a swiping gesture on the spherical display could be a way to control the timeline of animated data visualizations, by swiping right to move forward (future) and swiping left to move backward (past). Such functionality would allow precise control of time. Similarly, this gesture can also be used to navigate between data visualizations without needing a button or an external control system.

\paragraph{Split views.}

Another way to facilitate comparisons is to use a split-view, where the same globe slice displays side-by-side visualizations. It also allows multitasking and reduces conflicts by allowing independent interactions \cite{carapina_exploring_2023}. 


\paragraph{Pinned or floating overlay windows.}

An overlay is a visual elements displayed on top of an existing interface. It functions similarly to a superimposed sheet that provides additional data or a control panel without modifying the main display. Overlays can contextualize visualization with localized information (text/image/video) on the spherical display (see \autoref{fig:floating-overlay}). They can either be pinned to a specific geolocation, or floating (locked in physical space). Pinned overlay windows can deliver contextualized information tied to a specific geographic region, ensuring persistent relevance as the globe rotates. Conversely, floating overlay windows are suited for control panels, with their positioning unaffected by the globe’s orientation. 
Some of the standard features of overlay windows include minimizing, closing, resizing, and repositioning. Multiple overlay windows can be rearranged next to each other to facilitate comparative analysis across different areas of the spherical display. 
Some specific features such as sliders or media controls can be included in these overlay windows, to adjust the timeline of the animated data or the amount of data displayed through opacity. 
The participants also suggested a movable magnifying glass. This floating overlay window could continuously display a detailed, zoomed-in view of the area beneath it. 

\begin{figure}
\centering
\includegraphics[width=\linewidth]{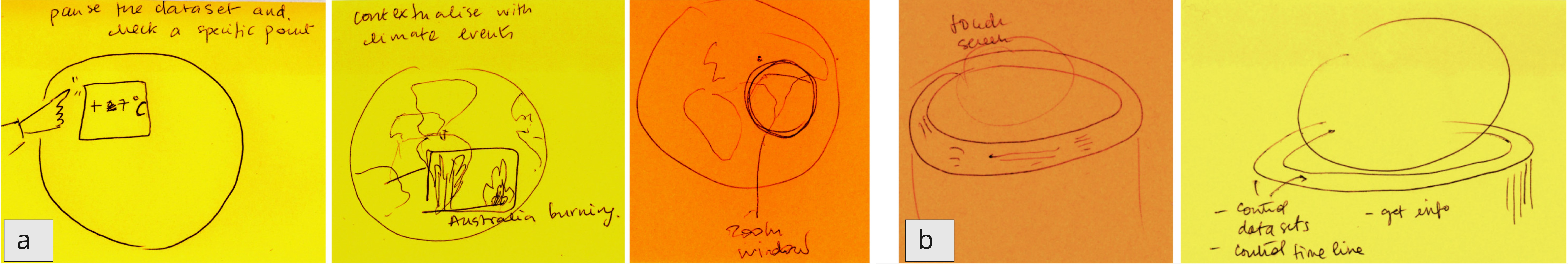}
\caption{Workshop sketches of (a) floating overlay windows and (b) fixed globe areas for additional displays.}
\label{fig:sticky-notes}
\end{figure}



\paragraph{Separate flat displays.}

Participants mentioned that additional displays can enhance understanding of data visualization by providing more space to control the amount of information, access the legend and data visualization description, and explore external information. In addition, they highlighted that the extra space could be used to navigate the data visualizations in different forms, including obtaining information for specific data points, controlling the timeline, changing the data visualizations, and comparing them. The discussions among the participants suggested that these additional displays could be implemented as handheld touch-enabled devices, custom made displays situated around the globe, or large projected displays (see \autoref{fig:sticky-notes}). Participants also mentioned that sonification \cite{enge_open_2024} and audio cues could be combined with additional displays to enhance the user experience and support understanding and navigation of data visualizations. 

\section{Conclusion and directions for future research}

This explorative study identifies research directions for an under-studied interaction device, namely spherical displays. We focused on the topic of climate discussions, as it is particularly relevant for, and suited to, the device. However, our findings can be applicable to the broader scope of learning and communication in a science center context \cite{besanccon2022exploring}. Through two workshops we identified interaction opportunities and challenges, which we map to implementation suggestions. Our results highlight the important potential of these devices to support climate conversations and be used in learning contexts that are getting increasingly important for visualization work (see e.g., \cite{Schonborn,Yu}). With this work, we aim to scaffold future work from the HCI and Visualization communities about these unconventional, yet communicatively salient, interactive devices.

\begin{acks}

We thank \ifanonymous\else \href{https://liu.se/medarbetare/matar63}{Mattias Arvola} \fi for his feedback throughout the study. This work was partially supported by \ifanonymous\else Marcus and Amalia Wallenberg Foundation (2023.0128) and (2023.0130), and Knut and Alice Wallenberg Foundation (2019.0024). \fi
\end{acks}

\bibliographystyle{ACM-Reference-Format}
\bibliography{references}

\end{document}